\documentclass[12pt]{article}
\usepackage{graphicx}
\usepackage[cp1251]{inputenc}
\usepackage{rotating}
\begin{document}
 \noindent {\footnotesize\it Astronomy Letters, 2017, Vol. 43, pp. 152--158}
 \newcommand{\dif}{\textrm{d}}

 \noindent
 \begin{tabular}{llllllllllllllllllllllllllllllllllllllllllllll}
 & & & & & & & & & & & & & & & & & & & & & & & & & & & & & & & & & & & & & \\\hline\hline
 \end{tabular}

 \bigskip
 \bigskip
 \centerline{\large\bf Kinematics of the Galaxy from Cepheids with proper}
 \centerline{\large\bf motions from the Gaia DR1 catalogue}
 \bigskip
 \bigskip
 \centerline{\bf V.V. Bobylev \footnote {e-mail: vbobylev@gao.spb.ru}}
 \bigskip
 \centerline {\small \it Central (Pulkovo) Astronomical Observatory, Russian Academy of Sciences}
 \bigskip
 \bigskip
 {
The sample of classic Cepheids with known distances and
line-of-sight velocities is supplemented by the proper motions
from the Gaia DR1 catalog. From spatial velocities of 260 stars
the components of the peculiar Solar velocity:
 $(U,V,W)_\odot=(7.90,11.73,7.39)\pm(0.65,0.77,0.62)$~km s$^{-1}$,
parameters of the Galactic rotation curve:
$\Omega_0 =28.84\pm0.33$~km s$^{-1}$ kpc$^{-1}$,
 $\Omega^{'}_0=-4.05\pm0.10$~km s$^{-1}$ kpc$^{-2}$,
 $\Omega^{''}_0=0.805\pm0.067$~km s$^{-1}$ kpc$^{-3}$ are obtained.
For the adopted Galactocentric Solar distance $R_0=8$~kpc the
linear circular velocity of the Local Standard of Rest is found as
$V_0=231\pm6$~km s$^{-1}$.
 }

\section*{INTRODUCTION}
Cepheids are the most important data source for studying structure
and kinematics of the Galaxy. Thanks to the reliably established
``period-luminosity'' relation, the distances to the Cepheids can
be determined with high accuracy (the error is of about 10\%)
(Berdnikov, et al., 2000; Sandage \& Tammann, 2006). Using
distances in combination with high-precision proper motions of
about 220 classic Cepheids from the HIPPARCOS catalog~(1997)  the
Galaxy rotation parameters (Feast \& Whitelock, 1997; Mel'nik, et
al.,~2015), the parameters of the spiral structure (Mel'nik, et
al.,~1999; Bobylev \& Bajkova, 2012; Dambis, et al., 2015) and the
parameters of the warp of the Galactic disk (Bobylev, 2013a;
2013b) were revised. From data on older Cepheids of II~type the
parameters of the central bulge and the distance to the Galactic
center (Majaess, et al., 2009) were re-determined.

On 14 September 2016 the Gaia DR1 catalog was published. This
catalog was created as a combination of Tycho-2 catalog (H{\o}g,
et al., 2000) with observational data of a space satellite Gaia
(Prusti, et al., 2016; Brown, et al., 2016) received during the
first year of operation. This version is designated as TGAS
(Tycho--Gaia Astrometric Solution)(Michalik, et al., 2015; Brown,
et al., 2016; Lindegren, et al., 2016). This catalog contains
about 2 million brightest stars (up to $\sim11.^m5$). Random
errors of the parameters included in the Gaia DR1 catalog  are
comparable or less than those given in the HIPPARCOS and Tycho-2
catalogs. An average error of the parallaxes is about 0.3 mas
(milliarcseconds). This means that the distances with errors of
about 10\% cover the Solar neighborhood with radius of about
300~pc. However, in case of Cepheids the situation may be more
favorable. In the opinion of Casertano, et al. (2016), who
performed the analysis of the distances of 212 nearby ($r<2$~kpc)
Cepheids from the Gaia DR1 catalog, the errors of their
trigonometric parallaxes are possibly overestimated by 20\%.

At present for studying the structure and kinematics of the Galaxy
at large distances from the Sun (3~kpc, and more) the use of the
distance scale of classic Cepheids based on the relation
``period--luminosity'' remains as a topical problem. But the
situation with proper motions is different. For the majority part
of the stars of the TGAS version, the average error of proper
motions is about 1 mas yr$^{-1}$ (milliarcseconds per year), but
for ($\sim$94000) stars, common with the HIPPARCOS catalog, this
error is much smaller, being about 0.06 mas/yr (Brown, et al.,
2016). So, the kinematic analysis of these stars with
high-precision proper motions is of great interest.

The aim of this work is the revision of the Galactic rotation
parameters on the basis of the Cepheids with the proper motions
from the Gaia DR1 catalog. For this we analyze both full spatial
velocities and velocities calculated from only proper motions.

 \section*{METHOD}\label{method}
From observations we know three projections of the stellar
velocity: the line-of-sight velocity $V_r$ as well as the two
velocities $V_l=4.74r\mu_l\cos b$ and $V_b=4.74r\mu_b$ directed
along the Galactic longitude $l$ and latitude $b$ and expressed in
km s$^{-1}$. Here, the coefficient 4.74 is the ratio of the number
of kilometers in an astronomical unit to the number of seconds in
a tropical year, and $r$ is the heliocentric distance of the star
$r$ in kpc. The proper motion components $\mu_l\cos b$ and $\mu_b$
are expressed in milliarcseconds per year (mas yr$^{-1}$). The
velocities $U,V,$ and $W$ directed along the rectangular Galactic
coordinate axes are calculated via the components $V_r,V_l,$ and
$V_b:$
 \begin{equation}
 \begin{array}{lll}
 U=V_r\cos l\cos b-V_l\sin l-V_b\cos l\sin b,\\
 V=V_r\sin l\cos b+V_l\cos l-V_b\sin l\sin b,\\
 W=V_r\sin b                +V_b\cos b,
 \label{UVW}
 \end{array}
 \end{equation}
where $U$ is directed from the Sun to the Galactic center, $V$ is
in the direction of Galactic rotation, and $W$ is directed toward
the north Galactic pole. We can find two velocities, $V_R$
directed radially away from the Galactic center and $V_{circ}$
orthogonal to it and pointing in the direction of Galactic
rotation, based on the following relations:
 \begin{equation}
 \begin{array}{lll}
  V_{circ}= U\sin \theta+(V_0+V)\cos \theta, \\
       V_R=-U\cos \theta+(V_0+V)\sin \theta,
 \label{VRVT}
 \end{array}
 \end{equation}
where the position angle $\theta$ is calculated as
$\tan\theta=y/(R_0-x),$ while $x$ and $y$ are the rectangular
Galactic coordinates of the star (velocities $U,V,W$ are directed along the respective axes
$x,y,z$).

To determine the parameters of the Galactic rotation curve, we use
the equations derived from Bottlinger’s formulas in which the
angular velocity $\Omega$ was expanded in a series to terms of the
second order of smallness in $r/R_0:$
\begin{equation}
 \begin{array}{lll}
 V_r=-U_\odot\cos b\cos l-V_\odot\cos b\sin l\\
 -W_\odot\sin b+R_0(R-R_0)\sin l\cos b\Omega^{'}_0\\
 +0.5R_0(R-R_0)^2\sin l\cos b\Omega^{''}_0,
 \label{EQ-1}
 \end{array}
 \end{equation}
 \begin{equation}
 \begin{array}{lll}
 V_l= U_\odot\sin l-V_\odot\cos l-r\Omega_0\cos b\\
 +(R-R_0)(R_0\cos l-r\cos b)\Omega^{'}_0\\
 +0.5(R-R_0)^2(R_0\cos l-r\cos b)\Omega^{''}_0,
 \label{EQ-2}
 \end{array}
 \end{equation}
 \begin{equation}
 \begin{array}{lll}
 V_b=U_\odot\cos l\sin b + V_\odot\sin l \sin b\\
 -W_\odot\cos b-R_0(R-R_0)\sin l\sin b\Omega^{'}_0\\
    -0.5R_0(R-R_0)^2\sin l\sin b\Omega^{''}_0,
 \label{EQ-3}
 \end{array}
 \end{equation}
where $R$ is the distance from the star to the Galactic rotation
axis,
  \begin{equation}
 R^2=r^2\cos^2 b-2R_0 r\cos b\cos l+R^2_0.
 \end{equation}
$\Omega_0$ is the angular velocity of Galactic rotation at the
Galactocentric Solar distance $R_0,$ the parameters
$\Omega^{\prime}_0$ and $\Omega^{\prime\prime}$ are the
corresponding derivatives of the angular velocity, and
$V_0=|R_0\Omega_0|.$

It is necessary to adopt a certain value of the distance $R_0$.
One of the most reliable estimates of this value
$R_0=8.28\pm0.29$~kpc obtained by Gillessen, et al. (2009) from
the analysis of the orbits of the stars, moving around the
supermassive black hole at the center of the Galaxy. From masers
with trigonometric parallaxes Reid, et al. (2014) found
$R_0=8.34\pm0.16$~kpc. From analysis of kinematics of the masers
Bobylev \& Bajkova (2014) estimated $R_0=8.3\pm0.2$~kpc, in work
by Bajkova \& Bobylev (2015) it is found $R_0=8.03\pm0.12$~kpc,
Rastorguev, et al.(2016) obtained $R_0=8.40\pm0.12$~kpc. Recent
analysis of the orbits of stars moving around the supermassive
black hole at the center of the Galaxy gave an estimate
$R_0=7.86\pm0.2$~kpc (Boehle, et al., 2016). In the present work
the adopted value $R_0=8.0\pm0.2$~kpc.

 \section*{DATA}\label{data}
In the present work the catalog of classic Cepheids, described by
Mel'nik, et al. (2015), is used as a basis. This catalog contains
674 stars with the estimates of their heliocentric distances. For
a large number of stars the system heliocentric line-of-sight
velocities and proper motions are specified. The distances to the
Cepheids were determined on the basis of the
``period--luminosity'' relation using the modern calibrations
obtained from the infrared photometric observational data. The
original version of the catalog contains the proper motion from
the HIPPARCOS catalog (1997), which were taken from the version
revised by van Leeuwen ~(2007) .

Note that in the process of the identification of the catalogs,
the indexes from the HIPPARCOS catalog were added to about a dozen
of the Cepheids from the catalog of Mel'nik, et al.~(2015). The
line-of-sight velocities were added to two dozens of the Cepheids
found using the database SIMBAD
~(http://simbad.u-strasbg.fr/simbad/).

In the catalog of Mel'nik, et al.~(2015) the distances are known
for all Cepheids. We identified 290 Cepheids with the proper
motions from the Gaia DR1 catalog. 249 stars of them have the
system line-of-sight velocities. All of this allows to calculate
full spatial velocities of the stars and analyze their
three-dimensional movement. To estimate the individual age of the
Cepheids ($t$) from the period of the pulsations ($P$) we used the
calibration of Efremov~(2003) $\log t=8.5-0.65 \log P,$ received
by him from the Cepheids belonging to the open star clusters of
the Large Magellanic Cloud.

As it is noted in the work of Mel'nik, et al.~(2015) the Cepheid
distance scale used in the catalog was built for the
Galactocentric distance of the Sun $R_0=7.1$~kpc. We have adopted
the value $R_0=8.0$~kpc. Therefore, for obtaining self-consistent
solutions the distances of all the Cepheids were multiplied by
factor of $8.0/7.1=1.127$.

\begin{figure}[t]{\begin{center}
   \includegraphics[width=0.99\textwidth]{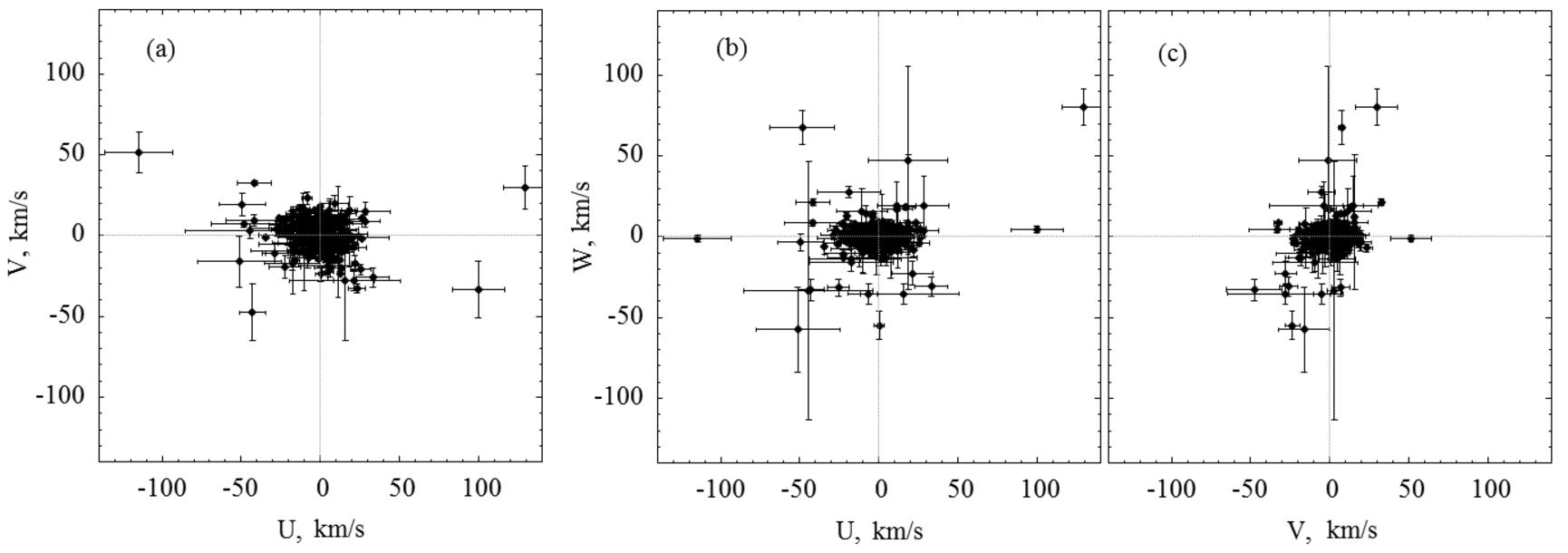}
 \caption{
The $UV$~a),$UW$~b) and  $VW$~c) residual spatial velocities of
the 249 Cepheids.
  }
  \label{f-UVW}
\end{center}}\end{figure}

 \section*{RESULTS AND DISCUSSION}\label{results}
The system of conditional equations ~(\ref{EQ-1})--(\ref{EQ-3})
was solved by the least squares method with the following weights:
 $w_r=S_0/\sqrt {S_0^2+\sigma^2_{V_r}},$
 $w_l=S_0/\sqrt {S_0^2+\sigma^2_{V_l}}$ and
 $w_b=S_0/\sqrt {S_0^2+\sigma^2_{V_b}},$
where $S_0$~is the ``cosmic'' variance,  $\sigma_{V_r},
\sigma_{V_l}, \sigma_{V_b}$~are the error variances of respective
observed velocities. The value of $S_0$ is comparable to the
standard root mean square residual $\sigma_0$ (unit weight error)
in the solution of conditional
equations~(\ref{EQ-1})--(\ref{EQ-3}). The value of this quantity
depends on the age of Cepheids and lies in the interval from
10~km/s for the youngest to 14~km/s for the oldest ones. In the
present work the value $S_0=14$~km/s is adopted. It was assumed
that the error of the distances to the Cepheids is 10\%.

A total number of the Cepheids with distances, line-of-sight
velocities and proper motions from the Gaia DR1 catalog is equal
to 249. In Fig.~\ref{f-UVW} the residual $UV$-, $UW$- and
$VW$-velocities of all 249 stars without any exceptions are shown.
From the figure we can see that there are a few Cepheids at high
velocities, especially on the $UW$ plane. These stars can spoil a
solution, so they should be excluded from the consideration. In
total there are 239 Cepheids, which satisfy the following
restrictions:
 \begin{equation}
 \begin{array}{lll}
  \sqrt{U^2+V^2+W^2}<120~\hbox {km s$^{-1}$},\\
                 |W|<60~\hbox {km s$^{-1}$},
 \label{criteri-1}
 \end{array}
\end{equation}
where the velocities $U,V,W$ are residual, they are exempt from
the differential rotation of the Galaxy using previously found
rotation parameters (or already known). Note that the equations
~(\ref{EQ-1})--(\ref{EQ-3}) we solved in two iteration with the
exception of stars with large residuals in accordance with the
$3\sigma$ criterion. In addition, we used 20 Cepheids for which
only the proper motion from the Gaia DR1 catalog are available.
These stars give the equations of the
form~(\ref{EQ-1})--(\ref{EQ-2}).

Solving the conditional equations (\ref{EQ-1})--(\ref{EQ-3}) for
260 Cepheids gave the following estimates for the kinematic
parameters:
 \begin{equation}
 \label{solution-333}
 \begin{array}{lll}
 (U_\odot,V_\odot,W_\odot)=\\
 (7.90,11.73,7.39)\pm(0.65,0.77,0.62)~\hbox{km s$^{-1}$},\\
      \Omega_0 =~28.84\pm0.33~\hbox{km s$^{-1}$ kpc$^{-1}$},\\
  \Omega^{'}_0 =-4.05\pm0.10~\hbox{km s$^{-1}$ kpc$^{-2}$},\\
 \Omega^{''}_0 =~0.805\pm0.067~\hbox{km s$^{-1}$ kpc$^{-3}$}.
 \end{array}
 \end{equation}
In this solution, unit weight error $\sigma_0=10.22$~km s$^{-1}$.
The rotation velocity of the Local Standard of Rest
$V_0=231\pm6$~km s$^{-1}$ (for an adopted value
$R_0=8.0\pm0.2$~kpc), and the Oort constants: $A=-16.20\pm0.38$~km
s$^{-1}$ kpc$^{-1}$ and $B= 12.64\pm0.51$~km s$^{-1}$ kpc$^{-1}$.
In table~\ref{t1} the kinematic parameters found by this method
for three samples of different age are given.

\begin{table}[t]\caption[]{\small
Kinematic parameters found for three samples of Cepheids of
different age groups using full spatial velocities
  }
\begin{center}      \label{t1}
\begin{tabular}{|l|r|r|r|r|r|}\hline
 Parameters & ${\overline t}=55$~Myr & ${\overline t}=95$~Myr & ${\overline t}=132$~Myr \\
                              &    $P>9^d$      &    $5^d<P\leq9^d$          &  $P<5^d$           \\\hline

 $U_\odot,$    km s$^{-1}$     & $ 5.7\pm1.2$  & $ 8.2\pm1.0$  & $ 9.9\pm1.2$ \\
 $V_\odot,$    km s$^{-1}$     & $13.8\pm1.6$  & $11.3\pm1.2$  & $11.2\pm1.5$ \\
 $W_\odot,$    km s$^{-1}$     & $ 6.6\pm1.2$  & $ 7.5\pm1.0$  & $ 7.9\pm1.1$ \\

$\Omega_0,$ km s$^{-1}$ kpc$^{-1}$     & $28.43\pm0.49$  & $28.85\pm0.55$  & $29.44\pm0.83$  \\
$\Omega^{'}_0,$ km s$^{-1}$ kpc$^{-2}$ & $-4.10\pm0.15$  & $-4.11\pm0.16$  & $-4.21\pm0.24$  \\
$\Omega^{''}_0,$ km s$^{-1}$ kpc$^{-3}$& $0.909\pm0.116$ & $0.795\pm0.106$ & $0.890\pm0.154$ \\
   $\sigma_0,$   km s$^{-1}$  &           10.2  &            9.9  &          10.6   \\
         $N_\star$            &            75   &             99  &            86   \\
   $A,$ km s$^{-1}$ kpc$^{-1}$  & $-16.40\pm0.58$ & $-16.45\pm0.63$ & $-16.84\pm0.97$ \\
   $B,$ km s$^{-1}$ kpc$^{-1}$   & $ 12.03\pm0.76$ & $ 12.40\pm0.83$ & $ 12.60\pm1.28$ \\
  \hline
\end{tabular}
\end{center}
\end{table}
\begin{figure}[t]
{\begin{center}
   \includegraphics[width=0.55\textwidth]{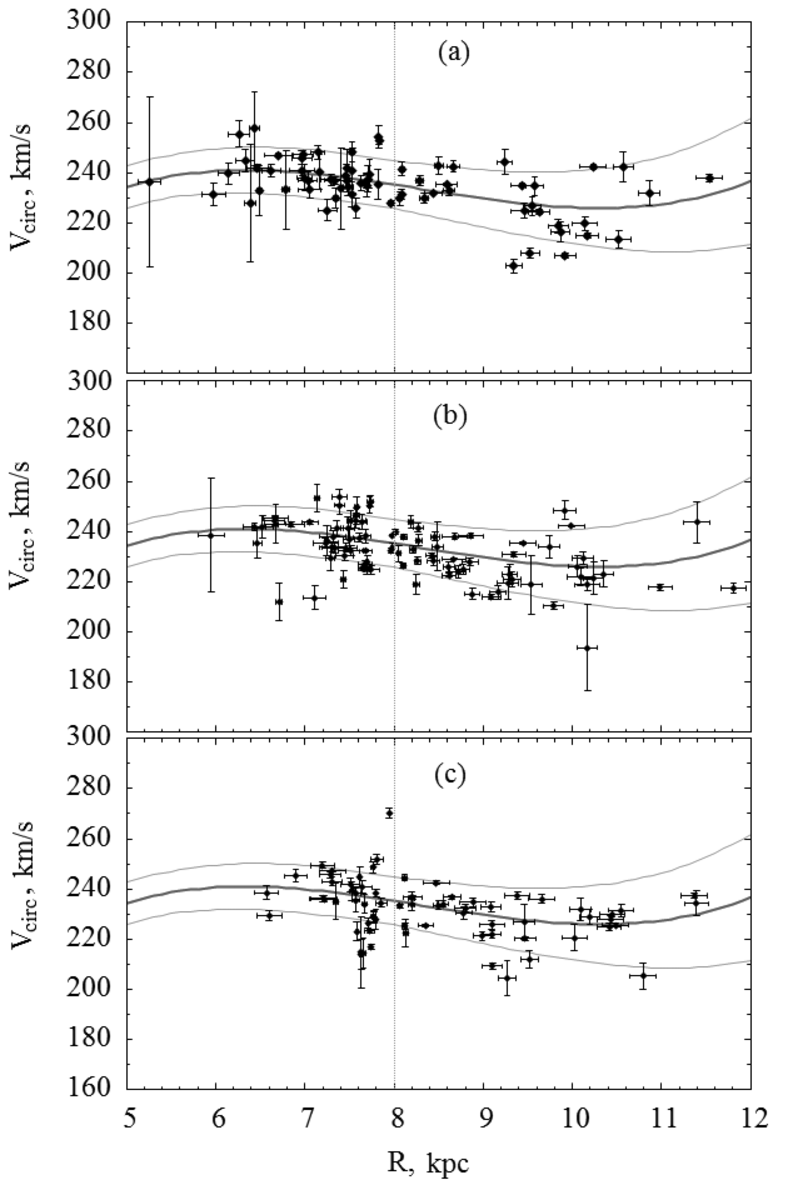}
 \caption{
Circular rotation velocities $V_{circ}$ versus distance $R$ (from
the Cepheid to the Galactic rotation axis) of young Cepheids~a),
middle-aged Cepheids ~b), old Cepheids~c); thick line represents
the rotation curve in accordance with the solution
~(\ref{solution-333}), thin lines mark the boundaries of the
confidence regions corresponding to a level of $1\sigma$.
  } \label{f-circ}
\end{center}}
\end{figure}
\begin{figure}[t]
{\begin{center}
   \includegraphics[width=0.55\textwidth]{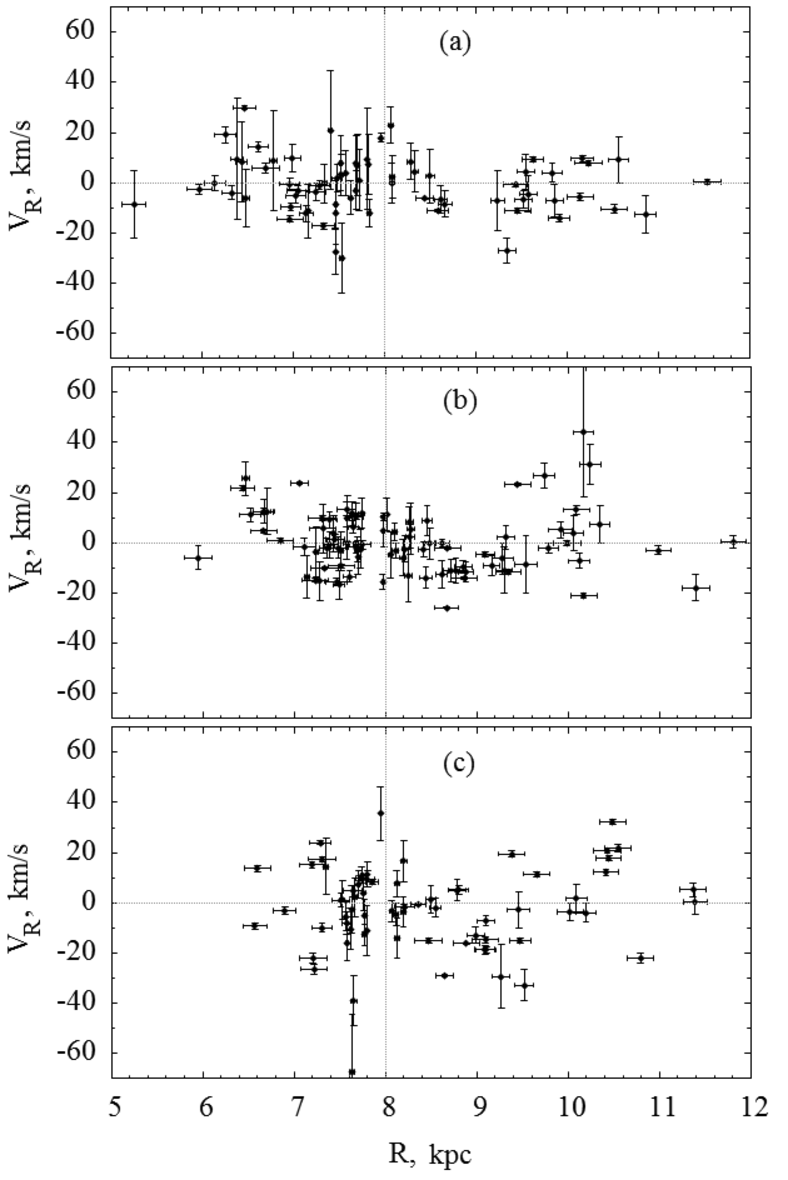}
 \caption{
Radial velocities $V_R,$ versus distance $R$ (from the Cepheid to
the Galactic rotation axis) of young Cepheids~a), middle-aged
Cepheids ~b), old Cepheids ~c).
  } \label{f-R}
\end{center}}
\end{figure}
\begin{figure}[t]
{\begin{center}
   \includegraphics[width=0.5\textwidth]{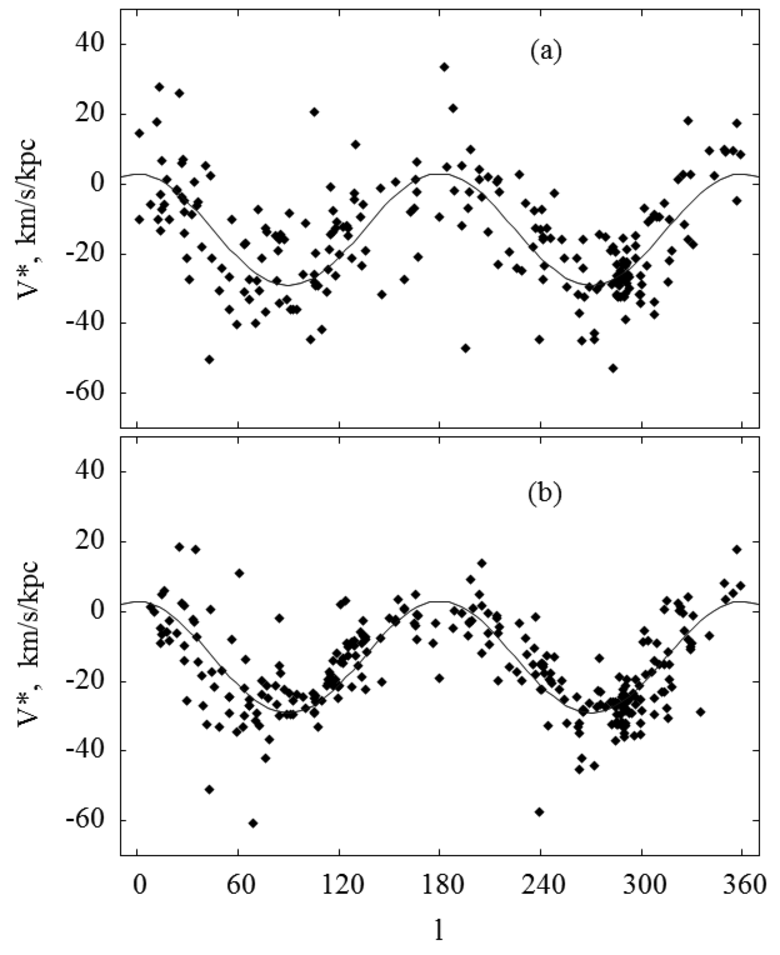}
 \caption{
Velocities $V^*$  versus Galactic longitude obtained from only
proper motions from the HIPPARCOS catalog a),  from the Gaia DR1
catalog b). The sine wave is drawn in accordance with the solution
~(\ref{solution-555}).
  } \label{f-mu-l}
\end{center}}
\end{figure}

In Fig.~\ref{f-circ} the  circular rotation velocities $V_{circ}$
versus distance $R$ for three samples of Cepheids of different age
groups are given. On each panel of this figure the same rotation
curve in accordance with the solution of~(\ref{solution-333}) is
shown. From the figure we can see that the Cepheids of different
age groups have small deviations from the specified rotation
curve, that is, they similarly take part in the Galactic rotation.
This conclusion is also confirmed by the closeness of the found
values of $\Omega_0,\Omega^{'}_0,\Omega^{''}_0$, specified in
table~\ref{t1}. On the other hand, in Fig.~\ref{f-circ} the
velocities $V_{circ}$ do not show visible periodic variations
caused by the Galactic spiral density wave.

In Fig.~\ref{f-R} the radial velocities $V_R,$ versus distance
$R$, for three samples of Cepheids of different age groups are
shown. In the velocities of young  Cepheids (panel a)), periodic
variations caused by the Galactic spiral density wave are clearly
visible. There is also a noticeable gradient of ages across the
Carina-Sagittarius spiral arm ($R\approx7$~kpc): old Cepheids are
located closer to the Solar circle, while the younger Cepheids are
farther from the Solar circle (closer to the center of the
Galaxy).

It is interesting to compare the proper motion from the HIPPARCOS
and Gaia DR1 catalogs. Following the Feast \& Whitelock (1997),
let's to rewrite equation~(\ref{EQ-2}) in the following way
 \begin{equation}
 \begin{array}{lll}
 V_l/r=(U_\odot\sin l-V_\odot\cos l)/r+\\
        \qquad+(2A\cos^2 l-\Omega_0)\cos b,
 \label{EQ-777}
 \end{array}
 \end{equation}
where $A=0.5\Omega'_0R_0$~is the Oort constant.

Moving components of the Solar peculiar velocity  to the left part
of equation (\ref{EQ-777}), we obtain, in the left part, the
velocity, freed from the Solar peculiar velocity, and in the right
part~--- members describing the rotation of a Galaxy:
 \begin{equation}
 \begin{array}{lll}
 V^*=[V_l-(U_\odot\sin l-V_\odot\cos l)]/r=\\
      \qquad =(2A\cos^2 l-\Omega_0)\cos b.
 \label{EQ-888}
 \end{array}
 \end{equation}
The velocity $V^*$ has the dimension of angular velocity, km
s$^{-1}$ kpc$^{-1}$. As it was shown by Feast \& Whitelock~(1997)
when analyzing Cepheids, the dependence of the velocity $V^*$ on
the longitude $l$ is a good illustration of the Galactic rotation.

In Fig.~\ref{f-mu-l} the velocities $V^*$ versus Galactic
longitude $l$ for each of the two proper motions, taken from the
HIPPARCOS and Gaia DR1 catalogues, are shown in panels a) and b)
respectively. In both panels the sine waves, reflecting the effect
of the rotation of the Galaxy, are perfectly visible. Meanwhile,
the points on panel b) have lower scatter relative to the sine
wave as compared to the points at panel a). So we can conclude
that the proper motions of the Cepheids from the Gaia DR1 catalog
are rather than the proper motions of the same stars from the
HIPPARCOS catalog.

In table~\ref{t2} the kinematic parameters, for three samples of
Cepheids of different age groups, found from only spatial
velocities $V_l$ are specified. These velocities were calculated
using the proper motions taken from the Gaia DR1 catalog. To find
the kinematic parameters the system of conditional equations
~(\ref{EQ-2}) with five unknowns was solved using the least
squares method. The Solar peculiar velocity component $W_\odot$
was fixed as $W_\odot=7$~km/s, because it can not be determined
only from equations of type ~(\ref{EQ-2}). Two
constraints~(\ref{criteri-1}) were used. From 260 Cepheids we
found the following estimates of the kinematic parameters:
 \begin{equation}
 \label{solution-555}
 \begin{array}{lll}
 (U_\odot,V_\odot)=(8.38,7.63)\pm(0.89,1.43)~\hbox{km s$^{-1}$},\\
      \Omega_0 =~29.04\pm0.71~\hbox{km s$^{-1}$ kpc$^{-1}$},\\
  \Omega^{'}_0 =-4.05\pm0.18~\hbox{km s$^{-1}$ kpc$^{-2}$},\\
 \Omega^{''}_0 =~0.778\pm0.117~\hbox{km s$^{-1}$ kpc$^{-3}$}.
 \end{array}
 \end{equation}
In this solution, unit weight error $\sigma_0=10.84$~km s$^{-1}$,
the Oort constants: $A=-16.20\pm0.71$~km s$^{-1}$ kpc$^{-1}$ and
$B= 12.84\pm1.00$~km s$^{-1}$ kpc$^{-1}$, the rotation velocity of
the Galaxy $V_0=232\pm8$~km/s for adopted value
$R_0=8.0\pm0.2$~kpc.

It is interesting to compare our solution ~(\ref{solution-555})
and the parameters listed in table~\ref{t2}, with the results of
Feast \& Whitelock~(1997) obtained using the same approach and the
proper motions from the HIPPARCOS catalog:
 $(U_\odot,V_\odot)=(9.3,11.2)\pm(1.5,1.5)$~km s$^{-1}$,
 $\Omega_0=27.19\pm0.87$~km s$^{-1}$ kpc$^{-1}$,
 $A=-14.82\pm0.84$~km s$^{-1}$ kpc$^{-1}$ and
 $B= 12.37\pm0.64$~km s$^{-1}$ kpc$^{-1}$.
We can see that the most noticeable differences there are in the
estimates of the velocities $U_\odot$ and especially $V_\odot$. On
the other hand, the results of Feast \& Whitelock~(1997) are in
good agreement with solution ~(\ref{solution-333}), but errors of
parameters $\Omega_0,$ $A$ and $B$ in ~(\ref{solution-333}) are
two times smaller.

From the analysis of the spatial velocities of 257 Cepheids with
proper motions from the HIPPARCOS catalog Mel'nik, et al.~(2015)
found the following solution:
 $(U_\odot,V_\odot)=(8.1,12.7)\pm(0.8,1.0)$~km s$^{-1}$,
 $\Omega_0=28.8\pm0.8$~km s$^{-1}$ kpc$^{-1}$,
 $\Omega^{'}_0=-4.88\pm0.14$~km s$^{-1}$ kpc$^{-2}$,
 $\Omega^{''}_0=1.07\pm0.17$~km s$^{-1}$ kpc$^{-3}$,
 $\sigma_0=10.84$~km s$^{-1}$ kpc$^{-1}$,
 $A=-18.3\pm0.6$~km s$^{-1}$ kpc$^{-1}$.
All these values are in good agreement with
solution~(\ref{solution-333}).

The most reliable estimates of the Galactic rotation parameters
were obtained by Rastorguev, et al. (2016) from data on 130 masers
with trigonometric parallaxes. For instance, the values for two
components of the peculiar velocity of the Sun are:
 $(U_\odot,V_\odot)=(11.40,17.23)\pm(1.33,1.09)$~km s$^{-1}$,
which are intrinsic for very young stars, and the following
parameters for the rotation curve of the Galaxy:
 $\Omega_0 =28.93\pm0.53$~km s$^{-1}$ kpc$^{-1}$,
 $\Omega^{'}_0=-3.96\pm0.07$~km s$^{-1}$ kpc$^{-2}$ and
 $\Omega^{''}_0=0.87\pm0.03$~km s$^{-1}$ kpc$^{-3}$,
the velocity of the Local Standard of Rest $V_0=243\pm10$~km
s$^{-1}$ for $R_0=8.40\pm0.12$~kpc. As it is seen, these results
and the solution ~(\ref{solution-333}) are in good agreement.

\begin{table}[t]\caption[]{\small
Kinematic parameters found for three samples of Cepheids of
different age groups using only velocities $V_l$
  }
\begin{center}      \label{t2}
\begin{tabular}{|l|r|r|r|r|r|}\hline
 Parameters   & ${\overline t}=55$~Myr & ${\overline t}=95$~Myr & ${\overline t}=132$~Myr \\
                              &    $P>9^d$      &    $5^d<P\leq9^d$          &  $P<5^d$           \\\hline

 $U_\odot,$    km s$^{-1}$    & $  6.0\pm1.9 $  & $ 7.9\pm1.4 $   & $11.2\pm1.5$ \\
 $V_\odot,$    km s$^{-1}$    & $ 11.5\pm3.4 $  & $ 7.9\pm2.3 $   & $ 5.8\pm2.2$ \\

     $\Omega_0,$ km s$^{-1}$ kpc$^{-1}$ & $ 28.1\pm1.4 $  & $ 28.4\pm1.1$   & $ 32.5\pm1.6$   \\
 $\Omega^{'}_0,$ km s$^{-1}$ kpc$^{-2}$ & $-4.11\pm0.32$  & $-4.23\pm0.28$  & $-4.01\pm0.34$  \\
$\Omega^{''}_0,$ km s$^{-1}$ kpc$^{-3}$ & $0.844\pm0.218$ & $0.970\pm0.191$ & $0.579\pm0.210$ \\
   $\sigma_0,$   km s$^{-1}$  &          10.0   &           10.2  &           9.8   \\
         $N_\star$            &            75   &             99  &            86   \\
             $A,$ km s$^{-1}$ kpc$^{-1}$ & $-16.4\pm1.3 $  & $-16.9\pm1.1 $  & $-16.0\pm1.4$ \\
             $B,$ km s$^{-1}$ kpc$^{-1}$ & $ 11.7\pm1.9 $  & $ 11.4\pm1.6 $  & $ 16.5\pm2.1$ \\
  \hline
\end{tabular}
\end{center}
\end{table}

 \section*{CONCLUSIONS}\label{conclusions}
The classic Cepheids with the proper motions contained in the Gaia
DR1 catalog are considered. We used the distances to the Cepheids
found by Mel'nik, et al.~(2015) on the basis of the
``period--luminosity'' relation using photometric data in infrared
band.

The sample of the 239 Cepheids for which it is possible to
calculate the spatial velocities was formed. In addition, 20 stars
without line-of-sight velocities, but with proper motions in the
Gaia DR1 catalog were used. From data of all these stars we found
components of the peculiar velocity of the Sun
 $(U_\odot,V_\odot,W_\odot)=(7.90,11.73,7.39)\pm(0.65,0.77,0.62)$~km s$^{-1}$
and the following parameters of the rotation curve of the Galaxy:
 $\Omega_0 =28.84\pm0.33$~km s$^{-1}$ kpc$^{-1}$,
 $\Omega^{'}_0=-4.05\pm0.10$~km s$^{-1}$ kpc$^{-2}$ and
 $\Omega^{''}_0=0.805\pm0.067$~km s$^{-1}$ kpc$^{-1}$ for the
 adopted value of the Galactocentric distance of the Sun
$R_0=8.0\pm0.2$~kpc. The circular velocity of the Local Standard
of Rest is $V_0=231\pm6$~km s$^{-1}$, the Oort constants
 $A=-16.20\pm0.38$~km s$^{-1}$ kpc$^{-1}$ and
 $B= 12.64\pm0.51$~km s$^{-1}$ kpc$^{-1}$.

This sample was divided into three groups depending on the age of
the Cepheids. It is shown that: a) there are no significant
differences from the age of the groups when we determine the
parameters of the Galactic rotation velocity $\Omega_0$,
$\Omega^{'}_0$ and $\Omega^{"}_0$, though previously it was
mentioned (Bobylev \& Bajkova, 2012; Mel'nik, et al.~2015) that
the most young Cepheids, by unknown reasons, rotate at a slower
velocity around the Galactic center; b) velocities $U_\odot$ and
$V_\odot$, determined from the sample of young Cepheids, differ
significantly from ones obtained from the samples of older
Cepheids.

There was also obtained the solution from only the proper motions
from the Gaia DR1 catalog:
 $(U_\odot,V_\odot)=(8.38,7.63)\pm(0.89,1.43)$~km s$^{-1}$,
      $\Omega_0=~29.04\pm0.71$~km s$^{-1}$ kpc$^{-1}$,
  $\Omega^{'}_0=-4.05\pm0.18$~km s$^{-1}$ kpc$^{-2}$,
 $\Omega^{''}_0=~0.778\pm0.117$~km s$^{-1}$ kpc$^{-3}$.
Here $V_0=232\pm8$~km s$^{-1}$ for $R_0=8.0\pm0.2$~kpc, the Oort
constants:
 $A=-16.20\pm0.71$~km s$^{-1}$ kpc$^{-1}$ and
 $B= 12.84\pm1.00$~km s$^{-1}$ kpc$^{-1}$.
In this case there is a difference between our estimates of
$U_\odot$ and $V_\odot$ and the estimates obtained by other
authors, but it does not exceeds the error level of $3\sigma$.

The astrometric observations from a space satellite Gaia are
continuing. The measurements of the star proper motions and
parallaxes will be significantly improved by the end of the
mission. Therefore, in the nearest future we can expect
significant refinement of the Galaxy kinematic parameters on the
basis of new Gaia data.

 \medskip
\subsection*{ACKNOWLEDGEMENTS}
The author is grateful to the referee for useful comments. This
work was supported by the ``Transient and Explosive Processes in
Astrophysics'' Program P--7 of the Presidium of the Russian
Academy of Sciences.
\subsection*{REFERENCES}

 1.~A.T. Bajkova and V.V. Bobylev, Baltic Astronomy {\bf 24}, 43 (2015).

 2. L.N. Berdnikov, A.K. Dambis, and O.V. Vozyakova,
    Astron. Astrophys. Suppl. {\bf 143}, 211, (2000).

 3.~V.V. Bobylev, A.T. Bajkova,  Astron. Lett. 38, 638 (2012). 

 4.~V.V. Bobylev, Astron. Lett. 39, 753 (2013). 

 5.~V.V. Bobylev, Astron. Lett. 39, 819 (2013). 

 6.~V.V. Bobylev, et al., Astron. Lett. 40, 389 (2014).   

 7.~A. Boehle, A.M. Ghez, R. Schodel, L. Meyer, S. Yelda, S. Albers,
    G.D. Martinez, E.E. Becklin, et al., arXiv: 160705726 (2016).

 8.~A.G.A. Brown, A. Vallenari, T. Prusti,J. de Bruijne, F. Mignard, R. Drimmel, et al., arXiv: 1609.04172, (2016).

 9.~S. Casertano, A.G. Riess, B. Bucciarelli, and M.G. Lattanzi,
     arXiv: 1609.05175 (2016).

 10.~A.K. Dambis, et al., Astron. Lett. 41, 489 (2015).

 11.~Yu.N. Efremov, Astron. Rep. 47, 1000 (2003).

 12.~M. Feast and P. Whitelock,
     Mon. Not. R. Astron. Soc. {\bf 291}, 683 (1997).

 13.~S. Gillessen, F. Eisenhauer, T.K. Fritz, H. Bartko, K. Dodds-Eden,
     O.~Pfuhl, T. Ott, and R. Genzel, Astroph. J. {\bf 707}, L114 (2009).

 14.~E.~H{\o}g, C.~Fabricius, V.V.~Makarov, U. Bastian, P. Schwekendiek,
     A. Wicenec, S. Urban, T. Corbin, and G. Wycoff),
     Astron. Astrophys. {\bf 355}, L~27 (2000). 

 15.~L. Lindegren, U. Lammers, U. Bastian, J. Hernandez, S. Klioner,
     D. Hobbs, A. Bombrun, D. Michalik, et al., arXiv: 1609.04303 (2016).

 16.~F. van Leeuwen, Astron. Astrophys. {\bf 474}, 653 (2007).

 17.~D.J. Majaess, D.G. Turner, and D.J. Lane,
     Mon. Not. R. Astron. Soc. {\bf 398}, 263 (2009).

 18.~A.M. Mel'nik, et al., Astron. Lett. 25, 518 (1999).

 19.~A.M. Mel'nik, P. Rautiainen, L.N. Berdnikov, A.K. Dambis,
     and A.S. Rastorguev, AN {\bf 336}, 70 (2015).

 20.~D. Michalik, L. Lindegren, and D. Hobbs, Astron. Astrophys. {\bf 574}, A115 (2015).

 21.~T. Prusti, J.H.J. de Bruijne, A.G.A. Brown,
     A. Vallenari, C. Babusiaux, C.A.L. Bailer-Jones, U. Bastian, M. Biermann, et al.,
     arXiv: 1609.04153, (2016).

 22.~A.S. Rastorguev,  M.V. Zabolotskikh, A.K. Dambis,
     N.D. Utkin, A.T. Bajkova, and V.V. Bobylev, arXiv: 1603.09124 (2016).

 23.~M.J. Reid, K.M. Menten, A. Brunthaler, X.W. Zheng, T.M. Dame,
     Y. Xu, Y.~Wu, B. Zhang, et al., Astrophys. J. {\bf 783}, 130 (2014). 

 24.~A. Sandage and G.A. Tammann,
     Ann. Rev. Astron. Astrophys. {\bf 44}, 93 (2006).

 25. The HIPPARCOS and Tycho Catalogues, ESA SP--1200 (1997).
\end{document}